\documentstyle[12pt,amssymb,amsmath]{article}
\newcommand{\p}{\partial}
\newcommand{\la}{\lambda}
\newcommand{\La}{\Lambda}

\begin{document}
\title{ {\bf Canonical explicit B\"{a}cklund transformations with spectrality 
for constrained flows of  soliton hierarchies}}
\author{ {\bf Yunbo Zeng\dag\hspace{1cm} Yiqing Zhu\dag \hspace{1cm}
Ting Xiao\dag\hspace{1cm}  Huihui Dai\ddag} \\
    {\small {\it\dag
 Department of Mathematical Sciences, Tsinghua University,
        Beijing 100084, China} }\\
            {\small {\it\ddag
        Department of Mathematics, City University of Hong Kong, Kowloon, Hong
        Kong, China}}}

\date{}
\maketitle

\renewcommand{\theequation}{\arabic{section}.\arabic{equation}}

{\bf Abstract } 
It is shown that  explicit B\"{a}cklund transformations (BTs) for 
the high-order constrained flows of soliton hierarchy can be constructed 
via their Darboux transformations and Lax representation, and these BTs 
are canonical transformations including B\"{a}cklund parameter $\eta$ and  
possess a spectrality property with respect to $\eta$ and  the 'conjugated' 
variable $\mu$ for which the pair $(\eta, \mu)$ lies on the spectral curve. 
As model we present the canonical explicit BTs with the spectrality for 
high-order constrained flows of the Kaup-Newell hierarchy and the KdV hierarchy.

\ \


\section{Introduction}
\setcounter{equation}{0}
\hskip\parindent
B\"{a}cklund transformations (BTs) are an important tool in  the studies of 
integrable systems \cite{1}. In recent years some new properties of BTs for 
finite-dimensional integrable Hamiltonian systems  have attracted some attention
(see \cite {2}-\cite {66}). These explicit BTs are  shown to be canonical 
transformations including B\"{a}cklund parameter $\eta$ and to satisfy 
a spectrality property with respect to $\eta$ and  the 'conjugated' variable 
$\mu$ for which the pair $(\eta, \mu)$ or $(\eta, f(\mu))$ for some function 
$f(\mu)$ lies on the spectral curve.
The spectrality property of BTs is connected with the problem of separation of variables, 
in fact, the sequence of B\"{a}cklund parameters $\eta_j$ together with the conjugate 
variables $\mu_j$ can be considered as the separation variables for the finite-dimensional 
integrable Hamiltonian systems . These BTs are defined as symplectic integrable maps and 
can be
viewed as time discretization of particular flows of Liouville integrable systems \cite{2,3,5,6}.

The constrained flows of soliton equations which can be transformed into
finite-dimensional Hamiltonian systems have been widely studied recently.  
 The Lax representation for the high-order constrained flows can always be 
deduced from the adjoint representation for the soliton hierarchy \cite{10,11}. 
We pointed out in \cite{111} that based on the results of Darboux transformations (DTs) 
for soliton hierarchy and the Lax representation for the constrained flows, 
one can  construct explicit  BTs including  B\"{a}cklund parameter $\eta$ for 
the high-order constrained flows of soliton equations. 
We proceed  further to develop the ideas with some new BTs and study the problem of 
constructing B\"{a}cklund transformation for the high-order constrained flows of 
the soliton hierarchy.
We  show  that these BTs  be canonical transformations by presenting their generating 
functions and possess the spectrality property with respect to $\eta$ and conjugate 
variable $\mu$. In contrast with the few example of this kind of BTs presented in 
\cite{2}-\cite{66}, this paper presents a way to find infinite number of explicit BTs 
with the properties mentioned above by means of DTs for the soliton hierarchy and the 
Lax representations for the high-order constrained flows.
We illustrate the idea by the high-order constrained flows of the Kaup-Newell hierarchy 
and KdV hierarchy.

The paper is organized as follows. In section 2, we briefly describe the high-order 
constrained flows of the Kaup-Newell hierarchy. In section 3 we first present  DTs for 
the constrained flows , then we construct infinite number of  explicit BTs for 
the high-order constrained flows of the Kaup-Newell hierarchy and show them to 
be canonical transformations and  to satisfy the spectrality property by using 
the first three high-order constrained flows as model. The infinite number of 
canonical explicit BTs with spectrality  for the constrained flows of the KdV hierarchy 
are presented in section 4 and 5.

\section{High-order constrained flows of the Kaup-Newell hierarchy}
\setcounter{equation}{0}
\hskip\parindent

The  Kaup-Newell hierarchy \cite{kaup}
\begin{equation}
\label{s3}
 u_{t_n}=\left( \begin{array}{c} 
 q \\ r \end{array} \right)_{t_n}
=J\left( \begin{array}{c} 
 c_{2n-1} \\ b_{2n-1} \end{array} \right)
=J\frac{\delta H_{2n-2}}{\delta u},
 \qquad n=0,1,\cdots,
\end{equation}
with
$$H_{2n}=\frac 1{2n}(4a_{2n+2}-qc_{2n+1}-rb_{2n+1}), \qquad
J=\left( \begin{array}{cc}
 0 & D \\ D&0\end{array} \right),$$
is associated with the following  Kaup-Newell spectral problem 
\begin{equation}
\label{s1}
 \psi_x=U(u,\la)\psi\equiv \left( \begin{array}{cc}
 -\lambda^2 & \la q \\\la r&\lambda^2\end{array} \right)\psi,\qquad
\psi=\left( \begin{array}{c} 
 \psi_1 \\ \psi_2 \end{array} \right),\qquad
u=\left( \begin{array}{c} 
 q \\ r \end{array} \right),
\end{equation}
and the evolution of $\psi$
\begin{equation}
\label{s2}
   \psi_{t_n}=
 V^{(n)}(u, \lambda)\psi\equiv
\sum_{i=0}^{n-1} \left( \begin{array}{cc}
 a_{2i}\la^{2n-2i} & b_{2i+1}\la^{2n-2i-1} \\ 
c_{2i+1}\la^{2n-2i-1} & -a_{2i}\la^{2n-2i} \end{array} \right)\psi,   
\end{equation}
where
$$a_0=1,\quad b_0=c_0=a_1=0, \quad b_1=-q,\quad c_1=-r$$
$$a_2=-\frac 12qr,\quad b_2=c_2=0,\quad b_3=\frac 12(q^2r+q_x),
\quad c_3=\frac 12(qr^2-r_{x}),...,$$
and in general, $a_{2m+1}=b_{2m}=c_{2m}=0$,
$$\left( \begin{array}{c} 
 c_{2m+1} \\ b_{2m+1} \end{array} \right)=L\left( \begin{array}{c} 
 c_{2m-1} \\ b_{2m-1} \end{array} \right)=-L^m\left( \begin{array}{c} 
 r \\ q \end{array} \right), \quad
 a_{m,x}=qc_{m+1}-rb_{m+1},$$
$$L=\frac 12\left( \begin{array}{cc}
 D-rD^{-1}qD &-rD^{-1}rD \\ -qD^{-1}qD&-D-qD^{-1}rD\end{array} \right),
\quad D=\frac {\partial}{\partial x},\quad DD^{-1}=D^{-1}D=1.$$

We have
$$\frac{\delta \lambda}{\delta q}=\phi_2^2,\qquad 
\frac{\delta \lambda}{\delta r}=-\phi_1^2.$$

The high-order constrained flows of the Kaup-Newell hierarchy consist of 
the equations obtained from the spectral problem (\ref{s1}) for $N$ distinct 
$\lambda_j$
and the restriction of the variational derivatives for conserved quantities 
$H_{2n}$ and $\lambda_j$ (see \cite{10,11,7,8,9})
\begin{subequations}
\label{s4}
\begin{equation}
\label{s4a}
\Phi_{1,x}=-\Lambda^2\Phi_{1}+q\La\Phi_{2},\qquad
\Phi_{2,x}=r\La\Phi_{}+\Lambda^2\Phi_{2},\qquad
 \end{equation}
\begin{equation}
\label{s4b}
\frac{\delta H_{2n}}{\delta u}-
 \frac 12\sum_{j=1}^N 
 \frac{\delta \lambda_j}{\delta u}
=\left( \begin{array}{c} 
 c_{2n+1} \\ b_{2n+1} \end{array} \right)
-\frac 12\left( \begin{array}{c} 
 <\Phi_2, \Phi_2> \\-<\Phi_1, \Phi_1> \end{array} \right)=0,
\end{equation}
\end{subequations}
where $n=0,1,\cdots, \Phi_{i}=(\phi_{i1},...,\phi_{iN})^T, i=1,2, 
\Lambda=\mbox{diag} (\lambda_1,...,\lambda_N), <.,.> $ denotes the inner product. 
The Lax representation for the constrained flow (\ref{s4}) is given by \cite{10,11}
\begin{equation}
\label{s5}
M^{(n)}_x=[U, M^{(n)}],
\end{equation}
with Lax matrix $M^{(n)}$
\begin{equation}
\label{s6}
 M^{(n)}=\left( \begin{array}{cc}
 A^{(n)} & B^{(n)} \\C^{(n)} & -A^{(n)}\end{array} \right) 
=V^{(n+1)}+M_0
\end{equation}
 $$M_0=\frac 12\sum_{j=1}^{N} \frac{1}{\lambda^2-\lambda_j^2}
 \left( \begin{array}{cc}
 \la^2\la_j^2\phi_{1j}\phi_{2j} & -\la\la_j^3{\phi_{1j}}^2 \\ 
 \la\la_j^3{\phi_{2j}}^2 & -\la^2\la_j^2\phi_{1j}\phi_{2j} 
 \end{array} \right),$$
and the Lax pair for (\ref{s4})
\begin{equation}
\label{s7}
 \psi_x=
 U(u,\lambda)\psi,
\qquad M^{(n)}(u, \lambda) \psi=\mu\psi.
\end{equation}
The spectral curve $\Gamma$,
$$\Gamma:
W(\la,\mu;\{P_i\})=det(M^{(n)}(u, \lambda)-\mu)=0$$
is
\begin{equation}
\label{s9}
W(\la,\mu;\{P_i\})=\mu^2-(A^{(n)}(\lambda))^2-B^{(n)}(\lambda)C^{(n)}(\lambda)=0,
\end{equation}
where $\{P_i\}$ are integrals of motion for (\ref{s4}).

We present the first three high-order constrained flows as follows.

(1) For $n=0$, (\ref{s4a}) becomes a finite-dimensional integrable Hamiltonian system (FDIHS)
 \begin{equation}
\label{s11}
\Phi_{1,x}=\frac {\p \widetilde H_0}{\p\Phi_2},\qquad
\Phi_{2,x}=-\frac {\p \widetilde H_0}{\p\Phi_1},
\end{equation}
$$\widetilde H_0=-<\La^2\Phi_1,\Phi_{2}>+\frac 14<\La\Phi_1,\Phi_{1}><\La\Phi_2,\Phi_{2}>,$$
with Lax matrix $M^{(0)}$
$$A^{(0)}=\la^2+ \frac 12\sum_{j=1}^N \frac{\la^2\la_j^2}{\lambda^2-\lambda_j^2}
 \phi_{1j}\phi_{2j}, \qquad
B^{(0)}=-\frac 12\lambda<\Lambda\Phi_{1}, \Phi_1>
-\frac 12\sum_{j=1}^N \frac{\la\la_j^3}{\lambda^2-\lambda_j^2}
 \phi_{1j}^2,$$
 \begin{equation}
\label{s12}
C^{(0)}=\frac 12\lambda<\Lambda\Phi_{2}, \Phi_2>
+ \frac 12\sum_{j=1}^N \frac{\la\la_j^3}{\lambda^2-\lambda_j^2}
 \phi_{2j}^2.
\end{equation}
The spectral curve $\Gamma$ is 
\begin{equation}
\label{s13}
W(\la, \mu; \{P_i\})=\mu^2-\la^4-\la^2P_{-1}-P_0
-\sum_{j=1}^N\frac{P_j}{\lambda^2-\lambda_j^2}=0,
\end{equation}
with $P_{-1}=-\widetilde H_0,$
$$P_0=<\La^4\Phi_1,\Phi_{2}>-\frac 14<\La^3\Phi_1,\Phi_{1}><\La\Phi_2,\Phi_{2}>$$
$$-\frac 14<\La\Phi_1,\Phi_{1}><\La^3\Phi_2,\Phi_{2}>+\frac 14<\La^2\Phi_1,\Phi_{2}>^2,$$
$$P_j=\la_j^6\phi_{1j}\phi_{2j}-\frac 14\la_j^5\phi^2_{2j}-\frac 14\la_j^5\phi^2_{1j}
+\frac 12\la_j^4\phi_{1j}\phi_{2j}<\La^2\Phi_1,\Phi_{2}>
-\frac 14\la_j^3\phi^2_{1j}<\La^3\Phi_2,\Phi_{2}>$$
$$+\frac 14\sum_{k\neq j} 
\frac{1}{\lambda_j^2-\lambda_k^2}(2\la_j^4\la_k^4\phi_{1j}\phi_{2j} \phi_{1k}\phi_{2k}
-\la_j^5\la_k^3\phi^2_{1k}\phi^2_{2j}-\la_j^3\la_k^5\phi^2_{1j}\phi^2_{2k}),\quad j=1,...,N.$$
$P_1,...,P_N$ are $N$ independent integrals of motion in involution for FDIHS (\ref{s11}).

(2) For $n=1$, (\ref{s4}) can be written as  a FDIHS
 \begin{equation}
\label{s14}
Q_{x}=\frac {\p \widetilde H_1}{\p P},\qquad
P_{x}=-\frac {\p \widetilde H_1}{\p Q},
\end{equation}
with
$$\widetilde H_1=-<\La^2\Phi_1,\Phi_{2}>-\frac {p_1}{\sqrt{2}} <\La\Phi_1,\Phi_{1}>
+\frac {q_1}{\sqrt{2}}<\La\Phi_2,\Phi_{2}>-q_1^2p_1^2,$$
$$Q=(\phi_{11},...,\phi_{1N},q_1)^T,\qquad 
P=(\phi_{21},...,\phi_{2N},p_1)^T,\qquad q_1=\frac {q}{\sqrt{2}}, 
\qquad p_1=\frac {r}{\sqrt{2}},$$
and Lax matrix $M^{(1)}$
$$A^{(1)}=\la^4-q_1p_1\la^2+ \frac 12\sum_{j=1}^N \frac{\la^2\la_j^2}{\lambda^2-\lambda_j^2}
 \phi_{1j}\phi_{2j},$$
$$B^{(1)}=-\sqrt{2}q_1\la^3-\frac 12\lambda<\Lambda\Phi_{1}, \Phi_1>
- \frac 12\sum_{j=1}^N \frac{\la\la_j^3}{\lambda^2-\lambda_j^2}
 \phi_{1j}^2,$$
 \begin{equation}
\label{s15}
C^{(1)}=-\sqrt{2}p_1\la^3+\frac 12\lambda<\Lambda\Phi_{2}, \Phi_2>
+ \frac 12\sum_{j=1}^N \frac{\la\la_j^3}{\lambda^2-\lambda_j^2}
 \phi_{2j}^2.
\end{equation}

(3) For $n=2$, by introducing the following Jacobi-Ostrogradsky coordinates
$$Q=(\phi_{11},...,\phi_{1N},q_1,q_2)^T,\qquad 
P=(\phi_{21},...,\phi_{2N},p_1,p_2)^T,$$
$$q_1=q,\qquad q_2=r,\qquad 
p_1=-\frac 3{16}qr^2+\frac 14r_x,\qquad p_2=\frac 3{16}q^2r+\frac 14 q_x,$$
(\ref{s4}) can be transformed into a FDIHS
 \begin{equation}
\label{s17}
Q_{x}=\frac {\p \widetilde H_2}{\p P},\qquad
P_{x}=-\frac {\p \widetilde H_2}{\p Q},
\end{equation}
with
$$\widetilde H_2=-<\La^2\Phi_1,\Phi_{2}>-\frac 12q_2<\La\Phi_1,\Phi_{1}>
+\frac 12q_1<\La\Phi_2,\Phi_{2}>$$ $$-\frac 1{64}q_1^3q_2^3-4p_1p_2
-\frac 3{4}q_1^2q_2p_1+\frac 3{4}q_1q_2^2p_2,$$
and Lax matrix $M^{(2)}$
$$A^{(2)}=\la^6-\frac 12q_1q_2\la^4+(p_2q_2-p_1q_1)\la^2
+ \frac 12\sum_{j=1}^N \frac{\la^2\la_j^2}{\lambda^2-\lambda_j^2}
 \phi_{1j}\phi_{2j},$$
$$B^{(2)}=-q_1\la^5+(\frac 18q_1^2q_2+2p_2)\la^3-\frac 12\la<\La\Phi_1,\Phi_1>
-\frac 12\sum_{j=1}^N \frac{\la\la_j^3}{\lambda^2-\lambda_j^2}
 \phi^2_{1j},$$
 \begin{equation}
\label{s18}
C^{(2)}=-q_2\la^5+(\frac 18q_1q_2^2-2p_1)\la^3+\frac 12\la<\La\Phi_2,\Phi_2>
+\frac 12 \sum_{j=1}^N \frac{\la\la_j^3}{\lambda^2-\lambda_j^2}
 \phi^2_{2j}.
\end{equation}
 
\section{BTs for high-order constrained flows of the Kaup-Newell hierarchy}
\setcounter{equation}{0}
\hskip\parindent

Let $\psi(x,\eta)$ be a solution of (\ref{s1}) and (\ref{s2}) with $\la=\eta.$ 
It is known \cite{1,li,zhang} that
 the Darboux transformation (DT) for the Kuap-Newell hierarchy is given by
\begin{equation}
\label{h1}
\bar\psi=T\psi, \qquad T= \left( \begin{array}{cc}
 f\lambda &-\eta \\-\eta&f^{-1}\la\end{array} \right), 
\qquad f=\frac {\psi_2(x,\eta)}{\psi_1(x,\eta)},
\end{equation}
and
 \begin{equation}
\label{h4}
\bar q=-2\eta f+qf^2, \qquad \bar r=\frac 2{f}\eta+\frac r{f^2},
\end{equation}
namely, $\bar \psi, \bar q$ and $\bar r$ satisfy (\ref{s1}), (\ref{s2}) 
and the equation (\ref{s3}).

Now let $\psi(x,\eta)$ be a solution of (\ref{s7}) with $\la=\eta.$ Motivated by 
the DT for the Kaup-Newell hierarchy, it can be shown that the DT for 
the constrained flows (\ref{s4}) consists of (\ref{h1}), (\ref{h4}) and 
$$\bar\phi_{1j}=\frac 1{\sqrt{\la_j^2-\eta^2}}(\la_j\phi_{1j}f-\eta\phi_{2j}),$$
\begin{equation}
\label{h14}
 \bar\phi_{2j}=\frac {1}{\sqrt{\la_j^2-\eta^2}}(f^{-1}\la_j\phi_{2j}-\eta\phi_{1j}),
\quad j=1,...,N,
\end{equation}
namely, under the transformation (\ref{h1}), (\ref{h4}) and (\ref{h14}), 
$\bar q,\bar r, \bar\psi$ and  $\bar\phi_{1j},\bar\phi_{2j}$ satisfy (\ref{s4}) and (\ref{s7}).
We have
\begin{equation}
\label{h13}
  M^{(n)}(\bar u,\bar\Psi_1, \bar\Psi_2, \la)T=TM^{(n)}(u,\Psi_1,\Psi_2, \la).
\end{equation}

It follows from (\ref{s7})
\begin{equation}
\label{h15}
f=\frac {\mu-A^{(n)}(\eta)}{B^{(n)}(\eta)}
=\frac {C^{(n)}(\eta)}{\mu+A^{(n)}(\eta)}.
\end{equation}
By substituting (\ref{h15}) into (\ref{h4}) and (\ref{h14}), 
we obtain infinite number ($n=0,1,...$) of  explicit  BT $B_{\eta}$ 
for the constrained flows (\ref{s4}) as follows
\begin{subequations}
\label{h16}
\begin{equation}
\label{h16a}
\bar q=q\left(\frac {\mu-A^{(n)}(\eta)}{B^{(n)}(\eta)}\right)^2
-2\eta\frac {\mu-A^{(n)}(\eta)}{B^{(n)}(\eta)},
\end{equation}
\begin{equation}
\label{h16b}
\bar r= r\left(\frac {B^{(n)}(\eta)}{\mu-A^{(n)}(\eta)}\right)^2+\frac {2\eta
B^{(n)}(\eta)}{\mu-A^{(n)}(\eta)},
\end{equation}
\begin{equation}
\label{h16c}
\bar\phi_{1j}=\frac 1{\sqrt{\la_j^2-\eta^2}}
\left(\frac {\mu-A^{(n)}(\eta)}{B^{(n)}(\eta)}\la_j\phi_{1j}-\eta\phi_{2j}\right),
\end{equation}
\begin{equation}
\label{h16d}
\bar\phi_{2j}=\frac {1}{\sqrt{\la_j^2-\eta^2}}\left(\frac {
B^{(n)}(\eta)}{\mu-A^{(n)}(\eta)}\la_j\phi_{2j}-\eta\phi_{1j}\right),
\qquad j=1,...,N.
\end{equation}
\end{subequations}

It is found from (\ref{h1}) and (\ref{h13})
\begin{subequations}
\label{h17}
\begin{equation}
\label{h17a}
(\la^2-\eta^2)\overline A^{(n)}(\la)=(\la^2+\eta^2)A^{(n)}(\la)+f\la\eta B^{(n)}(\la)
-\frac {\la\eta}{f} C^{(n)}(\la),
\end{equation}
\begin{equation}
\label{h17b}
(\la^2-\eta^2)\overline B^{(n)}(\la)=2f\la\eta A^{(n)}(\la)+f^2\la^2 B^{(n)}(\la)
-\eta^2 C^{(n)}(\la)
\end{equation}
\begin{equation}
\label{h17c}
(\la^2-\eta^2)\overline C^{(n)}(\la)=-\frac {2\eta \la}f A^{(n)}(\la)-
\eta^2B^{(n)}(\la)+\frac {\la^2}{f^2}C^{(n)}(\la).
\end{equation}
\end{subequations}

Using the first constrained flows as model, we now show  the BTs (\ref{h16}) 
to be  canonical transformations by presenting their generating functions and 
check spectrality property  with respect to the B\"{a}cklund parameter $\eta$ 
and  the 'conjugated' variable $\mu$. 

(1) For the first constrained flow, the FDIHS (\ref{s11}), by comparing 
the coefficients of $\la^3$ in  (\ref{h17b}) after substituting (\ref{s12}), one gets
\begin{equation}
\label{h18}
 f=\frac {2\eta+\sqrt{4\eta^2
+<\La \Phi_1, \Phi_1><\La\overline \Phi_1,\overline \Phi_1>}}{<\La \Phi_1, \Phi_1>}.
\end{equation}
Then we have from (\ref{h14})
\begin{subequations}
\begin{equation}
\label{h19a}
 \phi_{2j}=-\frac {\sqrt{\la_j^2-\eta^2} \bar\phi_{1j}}{\eta}
+\frac {\la_j\phi_{1j}\sqrt{4\eta^2
+<\La \Phi_1, \Phi_1><\La\overline \Phi_1,\overline \Phi_1>}}{\eta<\La \Phi_1, \Phi_1>}
+\frac {2\la_j\phi_{1j}}{<\La \Phi_1, \Phi_1>}=\frac {\p S^{(0)}}{\p \phi_{1j}},
\end{equation}
\begin{equation}
\label{h19b}
\overline\phi_{2j}=\frac {\sqrt{\la_j^2-\eta^2} \phi_{1j}}{\eta}
-\frac {\la_j\bar\phi_{1j}\sqrt{4\eta^2
+<\La \Phi_1, \Phi_1><\La\overline \Phi_1,\overline \Phi_1>}}
{\eta<\La \overline\Phi_1, \overline\Phi_1>}+\frac {2\la_j\bar\phi_{1j}}
{<\La \overline\Phi_1, \overline\Phi_1>}=-\frac {\p S^{(0)}}{\p \overline\phi_{2j}},
\end{equation}
\end{subequations}
where the generating function $S^{(0)}$ for the canonical transformation (\ref{h16}) is 
given by
$$S^{(0)}_{\eta}=2\text{ln}\left(2\eta-\sqrt{4\eta^2
+<\La \Phi_1, \Phi_1><\La\overline \Phi_1,\overline \Phi_1>}\right)
-2\text{ln}<\La \overline\Phi_1, \overline\Phi_1>$$
\begin{equation}
\label{h20}
+\frac 1{\eta}\sqrt{4\eta^2+<\La \Phi_1, \Phi_1><\La\overline \Phi_1,\overline \Phi_1>}
-<\Delta \Phi_1, \overline\Phi_1>,
\end{equation}
where $\Delta=\text{diag}\left(\frac 1{\eta}\sqrt{\la_1^2-\eta^2},...,
\frac 1{\eta}\sqrt{\la_N^2-\eta^2}\right)$.

Furthermore, it is found from (\ref{s12}), (\ref{h14}) and (\ref{h15})
$$\tilde \mu\equiv\frac {\p S^{(0)}_\eta}{\p \eta}=-\frac 1{\eta^2}
\sqrt{4\eta^2+<\La \Phi_1, \Phi_1><\La\overline \Phi_1,\overline \Phi_1>}
+\frac 1{\eta^2}\sum_{j=1}^N\frac 1{\sqrt{\la_j^2-\eta^2}}\la_j^2\phi_{1j}\overline\phi_{1j}$$
\begin{equation}
\label{h21}
=\frac {2}{\eta^3}\left(A^{(0)}(\eta)+fB^{(0)}(\eta)\right)=\frac {2}{\eta^3}\mu,
\end{equation}
which implies that the pair $(\eta,\frac 12\eta^3\tilde\mu)$ lies on the spectral 
curve $\Gamma$
$$W(\eta, \frac 12\eta^3\tilde\mu;\{P_i\})=0,$$
namely satisfies the spectrality property \cite{2}.
Consider the composition $B_{\eta_1...\eta_N}=B_{\eta_1}\circ...\circ  B_{\eta_N}$ 
of the B\"{a}cklund transformation $B_{\eta_i}$. Then the corresponding generating 
function $S^{(0)}_{\eta_1...\eta_N}$ becomes the generating function of the canonical 
transformation from $(\Phi_1,\Phi_2)$ to $\{(\eta_i,\tilde \mu_i)\}$ given by the equations
$$\phi_{2j}=\frac {\p S^{(0)}_{\eta_1...\eta_N}}{\p \phi_{1j}},\qquad
\tilde\mu_{j}=\frac {\p S^{(0)}_{\eta_1...\eta_N}}{\p \eta_{j}}.$$
The separation variables $\{(\eta_i,\tilde\mu_i)\}$ satisfy the separation equations 
given by the spectral curve (\ref{s13})
$${\tilde\mu_i}^2=\frac 4{\eta_i^6}[\eta_i^4+\eta_i^2P_{-1}+P_0
+\sum_{j=1}^N\frac{P_j}{\eta_i^2-\lambda_j^2}], \qquad i=1,...,N.$$

(2) For the second constrained flow, the  FDIHS (\ref{s14}), 
by comparing the coefficients of $\la^5$ and $\la^3$ in  (\ref{h17b}) 
after substituting (\ref{s15}), one gets
\begin{subequations}
\label{h22}
\begin{equation}
\label{h22a}
f=\frac {\eta+\sqrt{\eta^2+2q_1\bar q_1}}{\sqrt {2} q_1},
\end{equation}
\begin{equation}
\label{h22b}
p_1=\frac 1{2\sqrt {2}\eta(\eta-\sqrt{2}fq_1)}\left(f^2<\La \Phi_1, \Phi_1>
-<\La \overline\Phi_1, \overline\Phi_1>+2\sqrt{2}\eta^2\bar q_1\right).
\end{equation}
\end{subequations}
Then by substituting (\ref{h22}),  we have from  (\ref{h14}) and (\ref{h4})
\begin{subequations}
\label{h23}
\begin{equation}
\label{h23a}
 \phi_{2j}=-\frac {\sqrt{\la_j^2-\eta^2} \bar\phi_{1j}}{\eta}
+\frac {\la_j\phi_{1j}}{\sqrt {2}q_1}+\frac {\la_j\phi_{1j}
\sqrt{\eta^2+2q_1\bar q_1}}{\sqrt {2}\eta  q_1}=\frac {\p S^{(1)}}{\p \phi_{1j}},
\end{equation}
\begin{equation}
\label{h23b}
\overline\phi_{2j}=\frac {\sqrt{\la_j^2-\eta^2} \phi_{1j}}{\eta}
-\frac {\sqrt{2}\la_jq_1\bar\phi_{1j}}{\eta(\eta+\sqrt{\eta^2+2q_1\bar q_1})}
=-\frac {\p S^{(1)}}{\p \bar\phi_{1j}},
\end{equation}
$$p_1=-\frac {(\eta+\sqrt{\eta^2+2q_1\bar q_1})^2}{4\sqrt{2}\eta q_1^2
\sqrt{\eta^2+2q_1\bar q_1}}<\La\Phi_{1}, \Phi_1>
-\frac {\eta \bar q_1}{\sqrt{\eta^2+2q_1\bar q_1}}$$
\begin{equation}
\label{h23c}
+\frac 1{2\sqrt{2}\eta\sqrt{\eta^2+2q_1\bar q_1}}<\La\bar\Phi_{1},\bar \Phi_1>
=\frac {\p S^{(1)}}{\p q_1},
\end{equation}
$$\bar p_1=\frac {\sqrt{2} q_1^2}{2\eta\sqrt{\eta^2+2q_1\bar q_1}
(\eta+\sqrt{\eta^2+2q_1\bar q_1})^2}<\La\overline\Phi_{1},\overline \Phi_1>
+\frac {\eta q_1}{\sqrt{\eta^2+2q_1\bar q_1}}$$
\begin{equation}
\label{h23d}
-\frac {1}{2\sqrt{2}\eta \sqrt{\eta^2+2q_1\bar q_1}}<\La\Phi_{1}, \Phi_1>
=-\frac {\p S^{(1)}}{\p \bar q_1},
\end{equation}
\end{subequations}
where the generating function $S^{(1)}$ for the canonical transformation 
(\ref{h16}) is given by
$$S^{(1)}=\frac {\eta+\sqrt{\eta^2+2q_1\bar q_1}}{2\sqrt{2}\eta q_1}
<\La\Phi_{1}, \Phi_1>-\eta\sqrt{\eta^2+2q_1\bar q_1}$$
\begin{equation}
\label{h24}
+\frac {q_1}{\sqrt{2}\eta(\eta+\sqrt{\eta^2+2q_1\bar q_1})}<\La\bar\Phi_{1},\bar \Phi_1>
-<\Delta \Phi_1, \overline\Phi_1>,
\end{equation}
where $\Delta=\text{diag}\left(\frac 1{\eta}\sqrt{\la_1^2-\eta^2},...,
\frac 1{\eta}\sqrt{\la_N^2-\eta^2}\right)$.

By a direct calculation, one can check the spectrality property by means of  
(\ref{h23}) and (\ref{s15})
$$\frac {\p S^{(1)}}{\p \eta}=-\frac {\bar q_1}{\sqrt{2}\eta^2 
\sqrt{\eta^2+2q_1\bar q_1}}<\La\Phi_{1}, \Phi_1>-\frac {q_1}
{\sqrt{2}\eta^2\sqrt{\eta^2+2q_1\bar q_1}}<\La\bar\Phi_{1},\bar \Phi_1>$$
\begin{equation}
\label{h25}
+\frac 1{\eta^2}\sum_{j=1}^N\frac 1{\sqrt{\la_j^2-\eta^2}}\la_j^2\phi_{1j}\overline\phi_{1j}
-\frac {2\eta^2+2q_1\bar q_1}{\sqrt{\eta^2+2q_1\bar q_1}}
=\frac {2}{\eta^3}[A^{(1)}(\eta)+fB^{(1)}(\eta)]=\frac {2}{\eta^3}\mu.
\end{equation}

(3) For the third constrained flow, the  FDIHS (\ref{s17}), by comparing 
the coefficient of $\la^7$ in  (\ref{h17b}) after substituting (\ref{s18}), one gets

\begin{equation}
\label{h26a}
f=\frac {\eta+\sqrt{\eta^2+q_1\bar q_1}}{q_1},
\end{equation}
Then  we have from (\ref{h14}), (\ref{h17b}) and (\ref{h17c})
\begin{subequations}
\label{h27}
\begin{equation}
\label{h27a}
 \phi_{2j}=\frac {\la_jf\phi_{1j}}{\eta}
-\frac {\sqrt{\la_j^2-\eta^2}}{\eta}\bar\phi_{1j}=\frac {\p S^{(2)}}{\p \phi_{1j}},
\end{equation}

\begin{equation}
\label{h27b}
\overline\phi_{2j}=-\frac {\la_j\bar\phi_{1j}}{\eta f}
+\frac {\sqrt{\la_j^2-\eta^2}}{\eta}\phi_{1j}
=-\frac {\p S^{(1)}}{\p \bar\phi_{1j}},
\end{equation}
\begin{equation}
\label{h27c}
 q_{2}=\frac {4(f^2p_2-\bar p_2)}{\eta\sqrt{\eta^2+q_1\bar q_1}}
-\frac {\bar q_1(\sqrt{\eta^2+q_1\bar q_1}+3\eta)}{2\sqrt{\eta^2+q_1\bar q_1}}
=-\frac {\p S^{(2)}}{\p p_2},
\end{equation}
\begin{equation}
\label{h27d}
\bar q_{2}=\frac {4(f^2p_2-\bar p_2)}{\eta f^2\sqrt{\eta^2+q_1\bar q_1}}
+\frac {q_1(3\eta-\sqrt{\eta^2+q_1\bar q_1})}{2\sqrt{\eta^2+q_1\bar q_1}}
=\frac {\p S^{(2)}}{\p \bar p_2},
\end{equation}
$$p_{1}=\frac 1{\eta(\sqrt{\eta^2+q_1\bar q_1})^3}[f^3(3\sqrt{\eta^2+q_1\bar q_1}
-\eta)p_2^2-q_1\bar p_2^2-2\bar q_1p_2\bar p_2+\frac 14(6\eta^3+3\eta q_1\bar q_1$$ 
$$-2\eta^2\sqrt{\eta^2+q_1\bar q_1}-2q_1\bar q_1\sqrt{\eta^2+q_1\bar q_1})\eta\bar p_2
-\frac 34\eta^2\bar q_1^2p_2-\frac 3{64} \eta^2q_1\bar q_1^2(q_1\bar q_1+4\eta^2)$$
\begin{equation}
\label{h27e}
+\frac 14(<\La\overline \Phi_1, \overline\Phi_1>-f^2<\La\Phi_1, \Phi_1>)(\eta^2+q_1\bar q_1)]
=\frac {\p S^{(2)}}{\p q_1},
\end{equation}
$$\bar p_{1}=\frac 1{\eta(\sqrt{\eta^2+q_1\bar q_1})^3}[-f^{-3}
(3\sqrt{\eta^2+q_1\bar q_1}+\eta)\bar p_2^2+\bar q_1 p_2^2+2 q_1p_2\bar p_2
-\frac 14(6\eta^3+3\eta q_1\bar q_1$$
$$+2\eta^2\sqrt{\eta^2+q_1\bar q_1}+2q_1\bar q_1\sqrt{\eta^2+q_1\bar q_1})\eta p_2
+\frac 34\eta^2 q_1^2\bar p_2+\frac 3{64} \eta^2q_1^2\bar q_1(q_1\bar q_1+4\eta^2)$$
\begin{equation}
\label{h27f}
+\frac 14(f^{-2}<\La\overline \Phi_1, \overline\Phi_1>-<\La\Phi_1, \Phi_1>)
(\eta^2+q_1\bar q_1)]
=-\frac {\p S^{(2)}}{\p \bar q_1},
\end{equation}
\end{subequations}
where the generating function $S^{(2)}$ for the canonical transformation (\ref{h16}) 
is given by
$$S^{(2)}=\frac 1{\eta\sqrt{\eta^2+q_1\bar q_1}}[-\frac 9{32}\eta^6-2f^2p_2^2
-2f^{-2}\bar p_2^2+\frac 12 \eta\bar q_1p_2(\sqrt{\eta^2+q_1\bar q_1}+3\eta)$$
$$-\frac 12 \eta q_1\bar p_2
(\sqrt{\eta^2+q_1\bar q_1}-3\eta)+4p_2\bar p_2]
-\frac 1{32}\eta(\sqrt{\eta^2+q_1\bar q_1})^3-\frac 3{16}\eta^2\sqrt{\eta^2+q_1\bar q_1}$$
\begin{equation}
\label{h28}
+\frac 1{2\eta f}<\La\overline \Phi_1, \overline\Phi_1>+\frac f{2\eta}<\La\Phi_1, \Phi_1>
-<\Delta \Phi_1, \overline\Phi_1>.
\end{equation}
By a straightforward calculation, one can check the spectrality property 
by means of  (\ref{h27}) and (\ref{s18})
\begin{equation}
\label{h29}
\frac {\p S^{(2)}}{\p \eta}
=\frac {2}{\eta^3}[A^{(2)}(\eta)+fB^{(2)}(\eta)]=\frac {2}{\eta^3}\mu.
\end{equation}

In the exactly same way we can construct infinite number of explicit BTs 
for the high-order constrained flows (\ref{s4}) $(n=3,4,...)$ with the properties 
mentioned above.

\section{High-order constrained flows of the KdV hierarchy}
\setcounter{equation}{0}
\hskip\parindent

Consider the Schr\"{o}dinger eigenvalue problem \cite{Ablo}
\begin{equation}
\label{a1}
 \phi_{xx}+(\lambda+u) \phi =0,
\end{equation}
which can be  rewritten as  in the matrix form
\begin{equation}
\label{a2}
 \left( \begin{array}{c} 
 \phi_1 \\ \phi_2 \end{array} \right)_x=
 U \left( \begin{array}{c}
 \phi_1 \\ \phi_2 \end{array} \right),
 \qquad
 U = \left( \begin{array}{cc}
 0 & 1 \\ -\lambda-u & 0 \end{array} \right).
\end{equation}
Take
\begin{equation}
\label{a3}
 \left( \begin{array}{c} 
 \phi_1 \\ \phi_2 \end{array} \right)_{t_n}=
 V^{(n)}(u, \lambda) \left( \begin{array}{c}
 \phi_1 \\ \phi_2 \end{array} \right),
\end{equation}
where
$$
 V^{(n)}=\sum_{i=0}^{n+1} \left( \begin{array}{cc}
 a_i & b_i \\ c_i & -a_i \end{array} \right)\lambda^{n+1-i}+\left( \begin{array}{cc}
 0 &0 \\b_{n+2} &0 \end{array} \right),$$
$$a_0=b_0=0,\quad c_0=-1,\quad a_1=0, \quad b_1=1,\quad c_1=-\frac 12u,$$
$$a_2=\frac 14u_x,\quad b_2=-\frac 12u,\quad c_2=\frac 18(u_{xx}+u^2),...,$$
and in general for $k=1,2,...,$
\begin{equation}
\label{a4}
 a_k=-\frac{1}{2}b_{k,x},\quad
 b_{k+1}=Lb_k=-\frac{1}{2}L^{k-1}u,
 \quad c_k=-\frac{1}{2}b_{k,xx}-b_{k+1}-b_k u,
\end{equation}
with
$$ L=-\frac{1}{4}\p^2-u+\frac{1}{2}\p^{-1}u_x,
 \quad \p=\frac{\partial}{\partial x},
 \quad \p\p^{-1}=\p^{-1}\p =1. $$

Then the compatibility condition of Eqs (\ref{a2}) and (\ref{a3}) 
gives rise to the KdV hierarchy
\begin{equation}
\label{a5}
 u_{t_n}= \p\frac{\delta H_n}{\delta u}=
 -2b_{n+2,x}, \qquad n=0,1,\cdots,
\end{equation}
where $H_n=\frac {4b_{n+3}}{2n+3}$.
We have
\begin{equation}
\label{a6}
\frac{\delta \lambda}{\delta u}=\phi_1^2.
\end{equation}

The high-order constrained flows of the KdV hierarchy consist of the equations 
obtained from the spectral problem (\ref{a2}) for $N$ distinct $\lambda_j$ and 
the restriction of the variational derivatives for the conserved quantities $H_n$ 
and $\lambda_j$ \cite{7, cao, anto}

\begin{subequations}
\label{a7}
\begin{equation}
\label{a7a}
\Phi_{1,x}=\Phi_{2}, \qquad \Phi_{2,x}=-(\Lambda+u)\Phi_{1},
\end{equation}
\begin{equation}
\label{a7b}
 D\left[\frac{\delta H_n}{\delta u} -
 \sum_{j=1}^N 
 \frac{\delta \lambda_j}{\delta u}\right] \equiv
D\left[-2b_{n+2}-\sum_{j=1}^N 
 {\phi_j}^2\right]=0,
\end{equation}
\end{subequations}
The auxiliary linear problems associated with (\ref{a7}) are
\begin{equation}
\label{a8}
 {\left( \begin{array}{c}
 \psi_1 \\ \psi_2 \end{array} \right)}_x
 = U {\left( \begin{array}{c}
 \psi_1 \\ \psi_2 \end{array} \right)},\qquad 
{\left( \begin{array}{c}
 \psi_1 \\ \psi_2 \end{array} \right)}_{t_n}
 = M^{(n)} {\left( \begin{array}{c}
 \psi_1 \\ \psi_2 \end{array} \right)},  
\end{equation}
with
$$ M^{(n)} 
 = \left( \begin{array}{cc}
 A^{(n)}&B^{(n)} \\C^{(n)} & -A^{(n)} \end{array} \right) $$  
$$  = \sum_{k=0}^{n+1} \left( \begin{array}{cc}
 a_k & b_k \\ c_k & -a_k \end{array} \right) 
 \lambda^{n+1-k}   
  +\frac 12 \sum_{j=1}^N \frac{1}{\lambda-\lambda_j}
 \left( \begin{array}{cc}
 \phi_{1j}\phi_{2j} & -\phi^2_{1j} \\ 
 \phi^2_{2j} & -\phi_{1j}\phi_{2j} 
 \end{array} \right).$$
The spectral curve $\Gamma$ is given by the formula (\ref{s9}).
We present the first three high-order constrained flows as follows.

(1) For $n=0$,  (\ref{a7}) becomes a FDIHS
 \begin{equation}
\label{a11}
\Phi_{1,x}=\frac {\p \widetilde H_0}{\p\Phi_2},\qquad
\Phi_{2,x}=-\frac {\p \widetilde H_0}{\p\Phi_1},
\end{equation}
$$\widetilde H_0=\frac 12<\Phi_2,\Phi_{2}>+\frac 12<\La\Phi_1,\Phi_{1}>
+\frac 14<\Phi_1,\Phi_{1}>^2,$$
with Lax matrix $M^{(0)}$
$$A^{(0)}=\frac 12\sum_{j=1}^N \frac{\phi_{1j}\phi_{2j}}{\lambda-\lambda_j},
 \qquad
B^{(0)}=1-\frac 12\sum_{j=1}^N \frac{\phi_{1j}^2}{\lambda-\lambda_j},$$
 \begin{equation}
\label{a12}
C^{(0)}=-\la-\frac 12<\Phi_{1}, \Phi_1>
+ \frac 12\sum_{j=1}^N \frac{\phi_{2j}^2}{\lambda-\lambda_j}.
\end{equation}
The spectral curve $\Gamma$ is 
\begin{equation}
\label{a13}
W(\la, \mu; \{P_i\})=\mu^2+\la-\sum_{j=1}^N \frac{P_j}{\lambda-\lambda_j}=0,
\end{equation}
with
$$P_j=\frac 12(\phi_{2j}^2+\la_j\phi_{1j}^2
+\frac 12<\Phi_{1}, \Phi_1>\phi_{1j}^2)
+\frac 12\sum_{k\neq j} 
\frac{\phi_{1j}\phi_{2j} \phi_{1k}\phi_{2k}-\phi^2_{1k}\phi^2_{2j}}{\lambda_j-\lambda_k},
\quad j=1,...,N.$$
$P_1,...,P_N$ are $N$ independent integrals of motion in involution for FDIHS (\ref{a11}).

(2) For $n=1$, (\ref{a7}) can be written as  a FDIHS
 \begin{equation}
\label{a14}
Q_{x}=\frac {\p \widetilde H_1}{\p P},\qquad
P_{x}=-\frac {\p \widetilde H_1}{\p Q},
\end{equation}
with
$$\widetilde H_1=\frac 12 <\La\Phi_1,\Phi_{1}>+\frac 12 <\La\Phi_2,\Phi_{2}>
+\frac 12 q<\Phi_1,\Phi_{1}>+\frac 18q_1^3+4p_1^2,$$
$$Q=(\phi_{11},...,\phi_{1N},q_1)^T,\qquad 
P=(\phi_{21},...,\phi_{2N},p_1)^T,\qquad q_1=u, \quad p_1=\frac 18u_x$$
and Lax matrix $M^{(1)}$
$$A^{(1)}=2p_1+ \frac 12\sum_{j=1}^N \frac{\phi_{1j}\phi_{2j}}{\lambda-\lambda_j},\qquad
B^{(1)}=\la-\frac 12 q_1- \frac 12\sum_{j=1}^N \frac{\phi_{1j}^2}{\lambda-\lambda_j},$$
 \begin{equation}
\label{a15}
C^{(1)}=-\la^2-\frac 12q_1\la-\frac 12<\Phi_{1}, \Phi_1>-\frac 14 q_1^2
+ \frac 12\sum_{j=1}^N \frac{\phi_{2j}^2}{\lambda-\lambda_j}.
\end{equation}

(3) For $n=2$, by introducing the following Jacobi-Ostrogradsky coordinates
$$Q=(\phi_{11},...,\phi_{1N},q_1,q_2)^T,\qquad 
P=(\phi_{21},...,\phi_{2N},p_1,p_2)^T,$$
$$q_1=u,\qquad q_2=u_{xx}+5u^2,\qquad 
p_1=-\frac 1{32}u_{xxx},\qquad p_2=-\frac 1{32}u_x,$$
(\ref{a7}) can be transformed into a FDIHS
 \begin{equation}
\label{a17}
Q_{x}=\frac {\p \widetilde H_2}{\p P},\qquad
P_{x}=-\frac {\p \widetilde H_2}{\p Q},
\end{equation}
with
$$\widetilde H_2=\frac 12<\Phi_2,\Phi_{2}>+\frac 12<\La\Phi_1,\Phi_{1}>
+\frac 12q_1<\Phi_1,\Phi_{1}>$$
 $$-\frac 5{32}q_1^2q_2-32p_1p_2+\frac 5{16}q_1^4-160q_1p_2^2+\frac 1{64}q_2^2,$$
and Lax matrix $M^{(2)}$
$$A^{(2)}=-8p_2\la+2p_1+12q_1p_2
+ \frac 12\sum_{j=1}^N \frac{\phi_{1j}\phi_{2j}}{\lambda-\lambda_j},$$
$$B^{(2)}=\la^2-\frac 12 q_1\la+\frac 18q_2-\frac 14q_1^2
-\frac 12\sum_{j=1}^N \frac{\phi^2_{1j}}{\lambda-\lambda_j},$$
$$C^{(2)}=-\la^3-\frac 12q_1\la^2+(\frac 18q_2-\frac 12q_1^2)\la
-\frac 12<\Phi_1,\Phi_1>+\frac 18q_1q_2$$
 \begin{equation}
\label{a18}
-\frac 38q_1^3-64p_2^2+\frac 12 \sum_{j=1}^N \frac{\phi^2_{2j}}{\lambda-\lambda_j}.
\end{equation}
  
\section{ BTs for high-order constrained flows of the KdV  hierarchy}
\setcounter{equation}{0}
\hskip\parindent

Let $\psi(x,\eta)$ be a solution of (\ref{a2}) and (\ref{a3}) with $\la=\eta.$ 
It is known \cite{1,13} that
there is the  DT for the KdV hierarchy given by
\begin{equation}
\label{k1}
\bar\psi=T\psi, \qquad T= \left( \begin{array}{cc}
 -f&1 \\-\la+\eta+f^2&-f\end{array} \right), 
\qquad f=\frac {\psi_2(x,\eta)}{\psi_1(x,\eta)},
\end{equation}
and
 \begin{equation}
\label{k2}
\bar u=u+2\text{ln} \psi_1=-u-2\eta-2f^2,
\end{equation}
namely under the transformation (\ref{k1}) and (\ref{k2}), $\bar \psi, \bar u
$ satisfy (\ref{a2}), (\ref{a3}) and  (\ref{a5}).

We now consider the DTs for high-order constrained flows (\ref{a7}). 
  Let $\psi(x,\eta)$ be a solution of (\ref{a8}) with $\la=\eta.$ 
Motivated by the DT for the KdV hierarchy, we find that the DT for 
the constrained flows (\ref{a7}) consists of (\ref{k1}), (\ref{k2}) and 
\begin{equation}
\label{k3}
\bar\phi_{1j}=\frac 1{\sqrt{\la_j-\eta}}(\phi_{2j}-f\phi_{1j}),\quad
 \bar\phi_{2j}=\frac {1}{\sqrt{\la_j-\eta}}[(\eta-\la_j+f^2)\phi_{1j}-f\phi_{2j}],
\quad j=1,...,N,
\end{equation}
namely under the transformation (\ref{k1}), (\ref{k2}) and (\ref{k3}), $\bar \psi, \bar u
$ and $\bar\phi_{1j},\bar\phi_{2j}$ satisfy (\ref{a7}) and (\ref{a8}).

By substituting (\ref{h15}) into (\ref{k2}) and (\ref{k3}), 
we obtain infinite number ($n=0,1,...$) of the  explicit  BT $B_{\eta}$ 
for the constrained flows (\ref{a7}) as follows
\begin{subequations}
\label{k4}
\begin{equation}
\label{k4a}
\bar u=-u-2\eta-2\left(\frac {\mu-A^{(n)}(\eta)}{B^{(n)}(\eta)}\right)^2,
\end{equation}
$$\bar\phi_{1j}=\frac 1{\sqrt{\la_j-\eta}}
\left(-\frac {\mu-A^{(n)}(\eta)}{B^{(n)}(\eta)}\phi_{1j}+\phi_{2j}\right),$$
\begin{equation}
\label{k4b}
\bar\phi_{2j}=\frac {1}{\sqrt{\la_j-\eta}}
\left[\left(\eta-\la_j+(\frac {\mu-A^{(n)}(\eta)}{
B^{(n)}(\eta)})^2\right)\phi_{1j}-\frac {\mu-A^{(n)}(\eta)}{
B^{(n)}(\eta)}\phi_{2j}\right),
\qquad j=1,...,N.
\end{equation}
\end{subequations}

It is found from (\ref{h13})
\begin{subequations}
\label{k5}
\begin{equation}
\label{k5a}
(\la-\eta)\overline A^{(n)}(\la)=(-\la+\eta+2f^2)A^{(n)}(\la)+f(f^2+\eta-\la)B^{(n)}(\la)
-f C^{(n)}(\la),
\end{equation}
\begin{equation}
\label{k5b}
(\la-\eta)\overline B^{(n)}(\la)=2fA^{(n)}(\la)+f^2B^{(n)}(\la)-C^{(n)}(\la),
\end{equation}
\begin{equation}
\label{k5c}
(\la-\eta)\overline C^{(n)}(\la)=-2f(f^2+\eta- \la) A^{(n)}(\la)-
(f^2+\eta-\la)^2B^{(n)}(\la)+f^2C^{(n)}(\la).
\end{equation}
\end{subequations}

(1) For the first constrained flow, the FDIHS (\ref{a11}), using (\ref{a12}) and (\ref{k5b}), 
one gets
\begin{equation}
\label{k6}
 f=\sqrt{-\eta-\frac 12< \Phi_1, \Phi_1>-\frac 12<\overline \Phi_1,\overline \Phi_1>}.
\end{equation}
Then we have from (\ref{k3})
\begin{subequations}
\label{k7}
\begin{equation}
\label{k7a}
 \phi_{2j}=\sqrt{\la_j-\eta} \bar\phi_{1j}
+\phi_{1j}\sqrt{-\eta-\frac 12< \Phi_1, \Phi_1>-\frac 12<\overline \Phi_1,\overline \Phi_1>}
=\frac {\p S^{(0)}}{\p \phi_{1j}},
\end{equation}
\begin{equation}
\label{k7b}
\bar\phi_{2j}=-\sqrt{\la_j-\eta} \phi_{1j}
-\bar\phi_{1j}\sqrt{-\eta-\frac 12< \Phi_1, \Phi_1>-\frac 12<\overline \Phi_1,\overline \Phi_1>}
=-\frac {\p S^{(0)}}{\p \bar\phi_{1j}},
\end{equation}
\end{subequations}
where the generating function $S^{(0)}$ for the canonical transformation  (\ref{k4}) 
is given by
$$S^{(0)}=-\frac 23\left(-\eta-\frac 12< \Phi_1, \Phi_1>
-\frac 12<\overline \Phi_1,\overline \Phi_1>\right)^{\frac 32}
+\sum_{j=1}^N \sqrt{\la_j-\eta}\phi_{1j}\bar\phi_{1j}.$$

Furthermore, it is found 
$$\frac {\p S^{(0)}}{\p \eta}=\frac 12\sum_{j=1}^N\frac 1{\eta-\la_j}\phi_{1j}\phi_{2j}
-\frac 12\sum_{j=1}^N\frac f{\eta-\la_j}\phi^2_{1j}+f$$
\begin{equation}
\label{k9}
=A^{(0)}(\eta)+fB^{(0)}(\eta)=\mu,
\end{equation}
which implies that the pair $(\eta,\frac {\p S^{(0)}}{\p \eta})$ lies on 
the spectral curve $\Gamma$ (\ref{a13}), namely the canonical transformation  
satisfies the spectrality property.
(\ref{k7}) and (\ref{k9}) were also given in \cite{5} in a little different way. 
Consider the composition $B_{\eta_1...\eta_N}=B_{\eta_1}\circ...\circ  B_{\eta_N}$ 
of the B\"{a}cklund transformation $B_{\eta_i}$. Then the corresponding generating 
function $S^{(0)}_{\eta_1...\eta_N}$ becomes the generating function of 
the canonical transformation from $(\Phi_1,\Phi_2)$ to $(\eta, \mu)$ given by the equations
$$\phi_{2j}=\frac {\p S^{(0)}_{\eta_1...\eta_N}}{\p \phi_{1j}},\qquad
\mu_{j}=\frac {\p S^{(0)}_{\eta_1...\eta_N}}{\p \eta_{j}}.$$
The separation variables $\{(\eta_i,\mu_i)\}$ satisfy the separation equations 
given by the spectral curve (\ref{a13})
$$\mu_i^2=-\eta_i+\sum_{j=1}^N \frac{P_j}{\eta_i-\lambda_j}, \qquad i=1,...,N.$$

(2) For the second constrained flow, the  FDIHS (\ref{a14}), 
by comparing the coefficients of $\la$ and $\la^0$ in  (\ref{k5b}) 
after substituting (\ref{a15}), one gets
\begin{subequations}
\label{k10}
\begin{equation}
\label{k10a}
f=\sqrt{-\eta-\frac 12(q_1+\bar q_1)},
\end{equation}
\begin{equation}
\label{k10b}
p_1=\frac 1{8f}(\eta\bar q_1+f^2q_1-\frac 12q_1^2-<\Phi_1, \Phi_1>
-<\overline\Phi_1, \overline\Phi_1>).
\end{equation}
\end{subequations}
Then by substituting (\ref{k10a}),  we have from  (\ref{k2}), (\ref{k3}) and (\ref{k10b})
\begin{subequations}
\begin{equation}
\label{k11a}
 \phi_{2j}=\sqrt{\la_j-\eta} \bar\phi_{1j}
+\sqrt{-\eta-\frac 12(q_1+\bar q_1)}\phi_{1j}=\frac {\p S^{(1)}}{\p \phi_{1j}},
\end{equation}
\begin{equation}
\label{k11b}
\bar\phi_{2j}=-\sqrt{\la_j-\eta}\phi_{1j}
-\sqrt{-\eta-\frac 12(q_1+\bar q_1)}\bar\phi_{1j}
=-\frac {\p S^{(1)}}{\p \bar\phi_{1j}},
\end{equation}
\begin{equation}
\label{k11c}
p_1=-\frac {2q_1^2+2\eta q_1-2\eta\bar q_1+q_1\bar q_1+2<\Phi_{1}, \Phi_1>
+2<\bar\Phi_{1},\bar \Phi_1>}{16\sqrt{-\eta-\frac 12(q_1+\bar q_1)}}
=\frac {\p S^{(1)}}{\p q_1},
\end{equation}
\begin{equation}
\label{k11d}
\bar p_1=\frac {2\bar q_1^2-2\eta(q_1-\bar q_1)+ q_1\bar q_1+2<\Phi_{1}, \Phi_1>
+2<\bar\Phi_{1},\bar \Phi_1>}
{16\sqrt{-\eta-\frac 12(q_1+\bar q_1)}}=-\frac {\p S^{(1)}}{\p \bar q_1},
\end{equation}
\end{subequations}
where the generating function $S^{(1)}$ for the canonical transformation  (\ref{k4}) 
is given by
$$S^{(1)}=\sum_{j=1}^N\sqrt{\la_j-\eta}\phi_{1j}\bar\phi_{1j}
+\frac 14q_1^2\sqrt{-\eta-\frac 12(q_1+\bar q_1)}
+\frac 12\sqrt{-\eta-\frac 12(q_1+\bar q_1)}<\Phi_{1}, \Phi_1>$$
$$+\frac 12\sqrt{-\eta-\frac 12(q_1+\bar q_1)}<\overline\Phi_{1},\overline \Phi_1>
+\frac 12q_1\left(\sqrt{-\eta-\frac 12(q_1+\bar q_1)}\right)^3
$$
\begin{equation}
\label{k12}
+\frac 25\left(\sqrt{-\eta-\frac 12(q_1+\bar q_1)}\right)^5
-\frac 12\eta \bar q_1\sqrt{-\eta-\frac 12(q_1+\bar q_1)}.
\end{equation}
By a direct calculation, one can find the spectrality property 
$$\frac {\p S^{(1)}}{\p \eta}=
-\frac 12\sum_{j=1}^N\frac {\phi_{1j}\bar\phi_{1j}}{\la_j-\eta}
+\frac 12f\sum_{j=1}^N\frac {\phi^2_{1j}}{\la_j-\eta}
-\frac 1{4f}(<\Phi_{1},\Phi_1>+<\overline\Phi_{1},\overline \Phi_1>)$$
\begin{equation}
\label{k13}
+\frac 1{8f}[q_1\bar q_1-8\eta^2-2\eta(q_1+\bar q_1)]
=A^{(1)}(\eta)+fB^{(1)}(\eta)=\mu.
\end{equation}

(3) For the third constrained flow, the  FDIHS (\ref{a17}), 
by comparing the coefficient of $\la^2$ in  (\ref{k5b}) 
after substituting (\ref{a18}), one gets
\begin{equation}
\label{k14}
f=\sqrt{-\eta-\frac 12(q_1+\bar q_1)},
\end{equation}
Then by substituting (\ref{k14}),  we have from  (\ref{k2}), (\ref{k3}) and (\ref{k5})
\begin{subequations}
\begin{equation}
\label{k15a}
 \phi_{2j}=\sqrt{\la_j-\eta} \bar\phi_{1j}
+\sqrt{-\eta-\frac 12(q_1+\bar q_1)}\phi_{1j}=\frac {\p S^{(2)}}{\p \phi_{1j}},
\end{equation}
\begin{equation}
\label{k15b}
\bar\phi_{2j}=-\sqrt{\la_j-\eta}\phi_{1j}
-\sqrt{-\eta-\frac 12(q_1+\bar q_1)}\bar\phi_{1j}
=-\frac {\p S^{(2)}}{\p \bar\phi_{1j}},
\end{equation}
\begin{equation}
\label{k15c}
q_2=-\bar q_2+8f^4+24\eta f^2+4q_1f^2-128p_2f+16\eta^2+12\eta q_1+6q_1^2
=-\frac {\p S^{(2)}}{\p p_2},
\end{equation}
$$ p_1=-\frac 14f^5+\frac 18q_1f^3-\frac 34\eta f^3+4p_2f^2-\frac 12\eta^2f
+\frac 38\eta q_1f+\frac 1{32}\bar q_2f-10q_1p_2+\frac 1{4f}[\frac 38q_1^3$$
\begin{equation}
\label{k15d}
+\frac 32\eta q_1^2+2\eta^2q_1-\frac 18q_1\bar q_2-64p_2^2-\frac 18\eta \bar q_2
+\frac 14\eta\bar q_1^2
-\frac 12<\Phi_{1}, \Phi_1>-\frac 12<\bar\Phi_{1},\bar \Phi_1>]
=\frac {\p S^{(1)}}{\p  q_1},
\end{equation}
\begin{equation}
\label{k15e}
\bar p_2=-p_2-\frac 18\eta f-\frac 18q_1f-\frac 18f^3
=-\frac {\p S^{(2)}}{\p \bar q_2},
\end{equation}
$$\bar p_1=-\frac 94f^5-\frac {25}{8}q_1f^3-\frac {19}4\eta f^3-8p_2f^2
-\frac 52\eta^2f-\frac {31}8\eta q_1f+\frac 3{32}\bar q_2f-\frac 98q_1^2f$$
$$-2q_1p_2-12p_2\eta-\frac 1{4f}[\frac 38q_1^3+\frac 32\eta q_1^2+2\eta^2q_1
-\frac 18q_1\bar q_2$$
\begin{equation}
\label{k15f}
-64p_2^2-\frac 18\eta \bar q_2+\frac 14\eta\bar q_1^2
-\frac 12<\Phi_{1}, \Phi_1>-\frac 12<\bar\Phi_{1},\bar \Phi_1>]
=-\frac {\p S^{(1)}}{\p  \bar q_1},
\end{equation}
\end{subequations}
where the generating function $S^{(2)}$ for the canonical transformation  
(\ref{k4}) is given by
$$S^{(2)}=\sum_{j=1}^N\sqrt{\la_j-\eta}\phi_{1j}\bar\phi_{1j}
+\frac 12f(<\Phi_{1}, \Phi_1>+<\bar\Phi_{1},\bar \Phi_1>)+\frac 18 f^3\bar q_2$$ 
$$+\frac f4[\frac 12\eta \bar q_2+\frac 12q_1\bar q_2
+256p_2^2-\frac 5{14} q_1^3-\frac 1{14}q_1^2 \bar q_1-\frac 17\eta q_1^2
-\frac 47q_1 \bar q_1^2+\frac 57\eta q_1\bar q_1-\frac 27\eta^2 q_1$$  
\begin{equation}
\label{k16}
+\frac 9{14}\bar q_1^3
-\frac 17\eta \bar q_1^2-\frac 27\eta^2 \bar q_1+\frac 87\eta^3]+
p_2\bar q_2+4\eta p_2\bar q_1-6p_2q_1^2-2p_2q_1\bar q_1 -4\eta p_2q_1-2p_2\bar q_1^2.
\end{equation}
By a direct calculation, one can find the spectrality property 
$$\frac {\p S^{(2)}}{\p \eta}=
A^{(2)}(\eta)+fB^{(2)}(\eta)=\mu.$$

Similarly we can construct infinite number of explicit BTs for the high-order 
constrined flows (\ref{a7}) $(n=3,4,...)$ with the properties mentioned above.

\section{Conclusion}
\setcounter{equation}{0}
\hskip\parindent
In contrast with few examples of  B\"{a}cklund transformations  with the properties 
described above presented in  \cite{2,3,4,5,6},  we propose a way to 
construct infinite number of explicit  B\"{a}cklund transformations for 
high-order constrained flows of soliton hierarchy by means of the Darboux transformations 
for soliton equations and the Lax representation for the high-order constrained flows. 
By constructing the generating functions, it is shown that these BTs are 
canonical transformations including B\"{a}cklund parameter $\eta$  and  
a spectrality property holds with respect to the B\"{a}cklund parameter $\eta$ 
and the conjugate variable $\mu$ with
the pair $(\eta,\mu)$ lying on the spectral curve.
\ \
\section*{ Acknowledgment }
The work described in this paper was supported by a grant from CityU (project
No. 7001072) and 
the Special Funds for Chinese Major Basic Research Project "Nonlinear Science".

\end{document}